\documentclass[aps,twocolumn, superscriptaddress, svgnames]{revtex4-2}
\usepackage[T1]{fontenc} 
\showhyphens{resonances}
\usepackage{tikz,tikzscale, relsize, pgfplots, xcolor, amssymb, amsmath}
\usetikzlibrary{backgrounds, arrows, calc, decorations, positioning, decorations.pathmorphing,decorations.pathreplacing, 
	decorations.markings, fixedpointarithmetic, fit, patterns, decorations.text, backgrounds,matrix,plotmarks, spy}
\usepgfplotslibrary{fillbetween}
\pgfplotsset{compat=1.17}
\usetikzlibrary{external}
\begin{document}
\title{Maximal Anderson Localization and Suppression of Surface Plasmons in Two-Dimensional Random Au Networks}

\author{J. Schultz}
\email{j.schultz@ifw-dresden.de}
\affiliation{Leibniz Institute for Solid State and Materials Research Dresden, Helmholtzstra{\ss}e 20, 01069 Dresden, Germany}
\author{K. Hiekel}
\affiliation{Physical Chemistry, TU Dresden, Zellescher Weg 19, 01069 Dresden, Germany}
\author{P. Potapov}
\affiliation{Leibniz Institute for Solid State and Materials Research Dresden, Helmholtzstra{\ss}e 20, 01069 Dresden, Germany}
\author{R. A. R{\"o}mer}
\affiliation{Department of Physics, University of Warwick, Coventry, CV4 7AL, United Kingdom}
\author{P. Khavlyuk}
\affiliation{Physical Chemistry, TU Dresden, Zellescher Weg 19, 01069 Dresden, Germany}
\author{A. Eychm{\"u}ller}
\affiliation{Physical Chemistry, TU Dresden, Zellescher Weg 19, 01069 Dresden, Germany}
\author{A. Lubk}
\email{a.lubk@ifw-dresden.de}
\affiliation{Leibniz Institute for Solid State and Materials Research Dresden, Helmholtzstra{\ss}e 20, 01069 Dresden, Germany}
\affiliation{Institute of Solid State and Materials Physics, TU Dresden, Haeckelstra{\ss}e 3, 01069 Dresden, Germany}

\begin{abstract}
Two-dimensional random metal networks possess unique electrical and optical properties, such as almost total optical transparency and low sheet resistance, which are closely related to their disordered structure. Here we present a detailed experimental and theoretical investigation of their plasmonic properties, revealing Anderson (disorder-driven) localized surface plasmon (LSP) resonances of very large quality factors and spatial localization close to the theoretical maximum, which couple to electromagnetic waves. Moreover, they disappear above a geometry-dependent threshold of ca. $1.7$\,eV in the investigated Au networks, explaining their large transparencies in the optical spectrum.
\end{abstract}

\maketitle
    \section{Introduction}
	
    Disordered nanostructures of noble metals, especially Au, are studied for their unique disorder-induced mechanical, electric, chemical and optical properties, and ensuing applications such as flexible electrodes for neural implants\,\cite{Minev2015} and monitoring of blood vessels\,\cite{Liu2017}, electromechanical chemical vapor and gas sensors\,\cite{Schlicke2017}, etc.\,\cite{Kim2019} One of their most intriguing properties is their ability to sustain strongly enhanced and localized electromagnetic fields coupled to localized surface plasmons (LSPs). These were first observed in so-called random dielectric thin films (also referred to as semicontinuous metal films)\,\cite{Brouers1997,Gresillon1999,Sarychev1999}, which occur upon Volmer-Weber-type growth of metallic thin films. Following the Ioffe-Regel criterion\,\cite{Ioffe1960}, Anderson localization (AL) of (classical and quantum) waves in disordered systems can occur in the limit of strong resonant scattering when the mean free path becomes equal or less than the wavelength. Accordingly, various articles discuss the observed emergence of localized surface plasmon modes in disordered systems in the context of AL employing different localization measures and theoretical frameworks (for quasistatic description see\,\cite{semicont_metal_film2007} and references therein, for fully retarded see\,\cite{Markel2006,Lubatsch2005}). Average quantities such as energy transport or effective dielectric functions may be well described with the help of renormalization techniques and sophisticated self-consistent theory of AL\,\cite{Lubatsch2005}. We note, that persisting theoretical challenges are met in the description of anisotropic and lossy media, and in the description of the local fluc\-tua\-tions, which may be ultimately traced to the vectorial and non-hermitian character of the plasmon dynamics. 
    
    Meanwhile, AL of classical optical and infrared fields has been observed in a variety of related systems, such as nanoparticle (NP) aggregates\,\cite{Abdellatif2016} and lithographically produced cavities\,\cite{Sapienza2010, Zhu2020}. Moreover, various applications such as Surface Enhanced Raman Spectroscopy (SERS)\,\cite{Brouers1997,Drachev2004}, the effective generation of non-linear optics on the nanoscale (e.g., four wave mixing, non-linear absorption, harmonic generation)\,\cite{Kauranen2012}, random la\-sing\,\cite{Cao_2005,Luan2015}, and bistable optical transistor\,\cite{Bergmann1994}, have been discussed and partially realized. Here, persisting challenges in engineering the network geometry  and the limited substrate choice of the semicontinuous metal films still hamper these efforts.\\	
    Very recently,  the development of a novel synthesis route allowed for the synthesis of large scale 2D Au networks with significantly reduced mass thickness (compared to the semicontinuous metal films), deliberately tunable coverage, and self-similar (fractal) character \,\cite{Hiekel2020}, thereby overcoming some of the above limitations. In the following, we experimentally demonstrate the emergence of disorder-driven LSP resonances in this novel type of Au networks. We demonstrate very large quality factors of $\approx8$ of these modes and show that the localization mecha\-nism can be described in terms of AL by comparison with simulations allowing to deliberately switch on loss, retardation, and disorder. For their characterization, we employ high-spatial resolution Scanning Transmission Electron Microscopy Electron Energy-Loss Spectroscopy (STEM-EELS) plasmon mapping\,\cite{Nelayah2007}, which permits to resolve the LSPs at nanometer length scale and a direct correlation of the localization position with the underlying Au network structure. 
	\section{Experiment}
	Two-dimensional Au networks of varying fractal dimension and coverage in the range between 31\% and 54\% have been synthesized following the procedures described in Ref.\,\cite{Hiekel2020}. Following the initial synthesis step and after the evaporation of the organic solvent the networks were transferred to a TEM-grid by carefully pressing the substrate onto the aqueous solution. Then, the substrate was washed with ethanol (EtOH) (see Fig.\,S1 in the supplement for an exemplary network).
	
\begin{figure}
		\centering
		\includegraphics{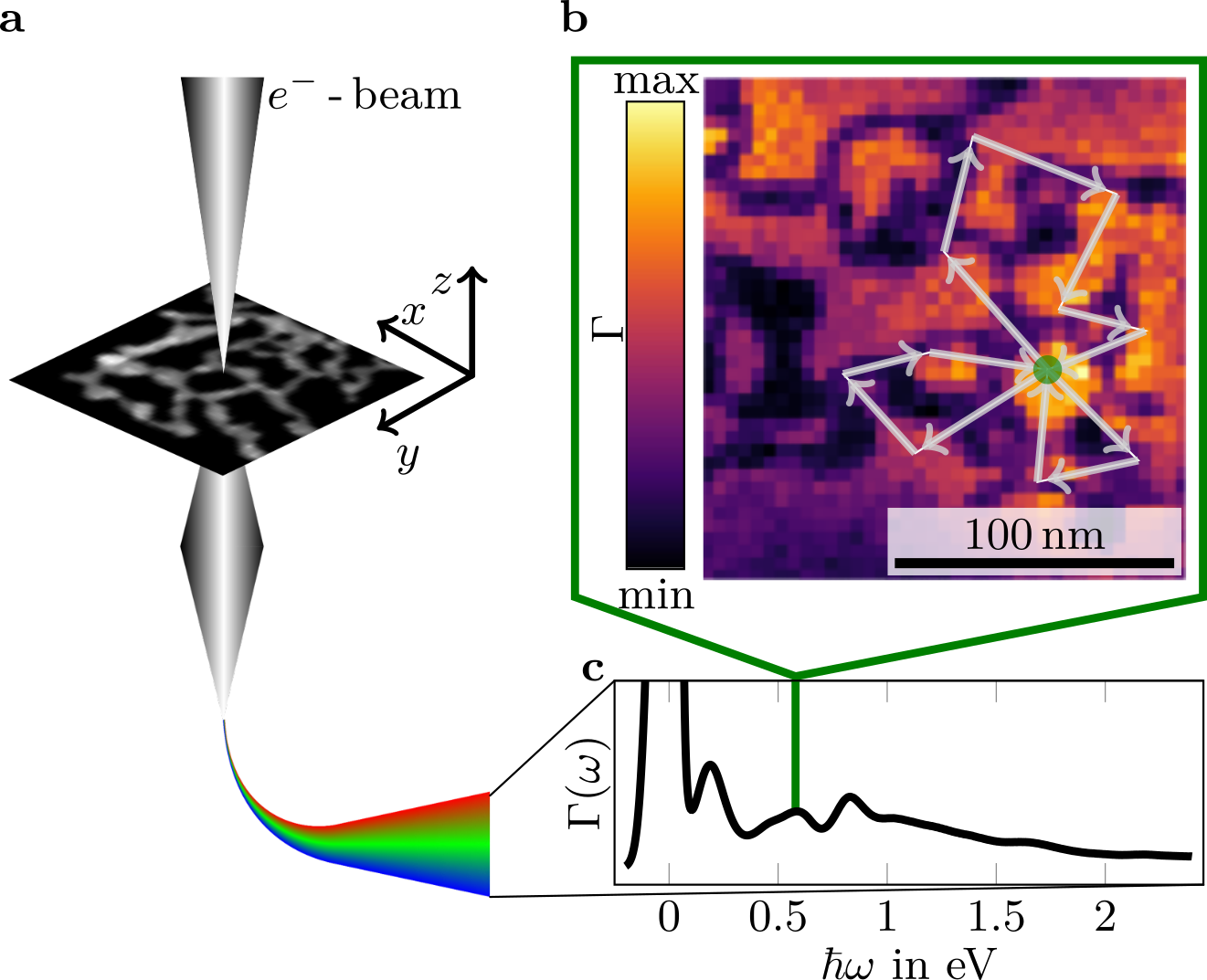}
		\caption{(a) Experimental setup. (b) 2D slice of the 3D dataset ($\Gamma(x,y,\omega)$) at $\hbar\omega=0.6$\,eV. The color scale corresponds to the spatially resolved loss probability, the gray arrows illustrate propagators of the surface plasmons eventually interferring constructively at random hot spots. (c) Spectrally resolved loss probability at a specific scan position.}
		\label{fig:setup}
	\end{figure}

\begin{figure*}
    \centering
		\includegraphics{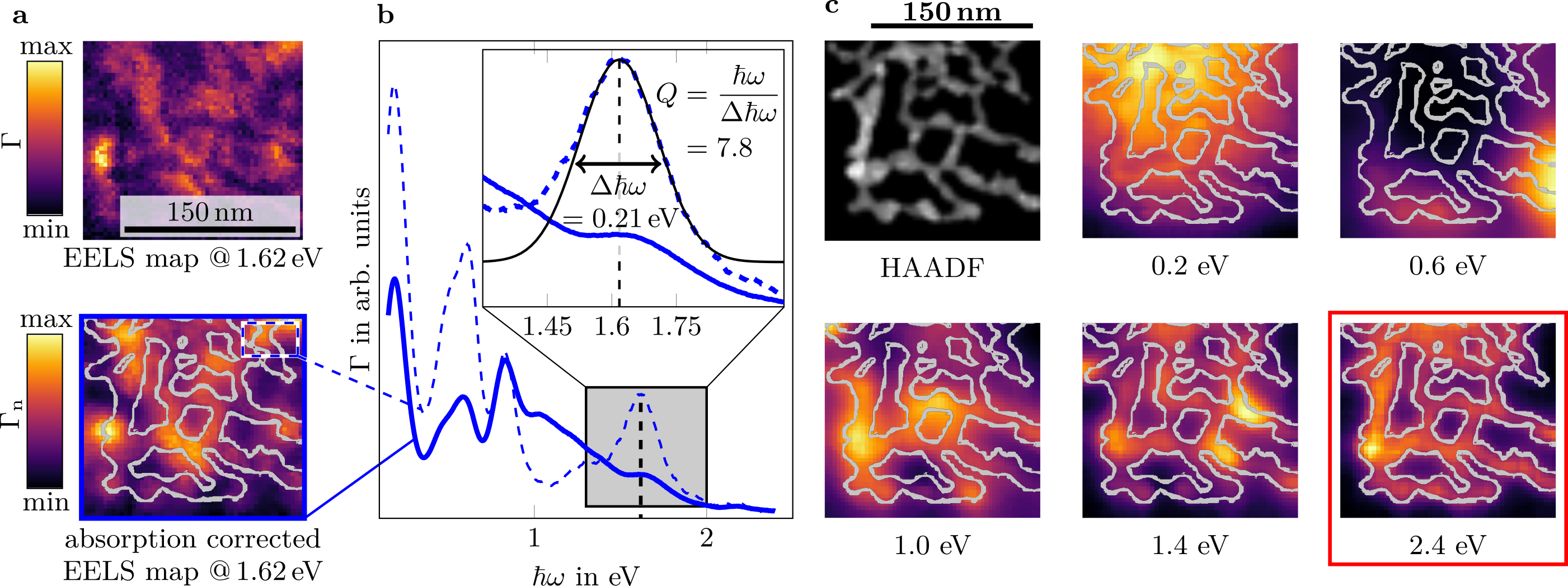}
		\caption{(a) As-recorded and absorption corrected loss probability map at $1.62$\,eV. (b) EELS spectra from different scanning regions indicated by the solid and dashed blue rectangle in (a). (c) High Angle Annular Dark Field (HAADF) image of EELS scanning region (a), and corresponding absorption corrected loss probability maps at different energies (the contours of the metal web are highlighted for better visibility). The intensity scaling of the absorption corrected loss probability maps is individual to compensate the effect of decreasing loss probability with increasing energy for better visibility of the hotspots.}
		\label{fig:spect_maps}
	\end{figure*}

	Spatially resolved EEL spectra (so-called spectrum images) of a collection of subareas of the network were recorded scanning a sharply focused probe of electrons with $80$\,keV kinetic energy over the sample (see Fig.\,\ref{fig:setup}) in a Transmission Electron Microscope (TEM). The utilized TEM is equipped with a probe corrector and a Wien-type monochromator improving the spectral reso\-lu\-tion to $60\,$-$\,70$\,meV. The recorded loss probability $\Gamma\left(\boldsymbol{r}_\perp=\left(x,y\right),\omega\right)$ corresponds to a projection of the induced $z$-component of the electric field along the beam direction $\left( \Gamma\left(\boldsymbol{r}_{\bot}, \omega\right)\propto\int\text{d}z\,\Re{\left\{e^{-i\omega z / v_{\text{z}}}\tilde{E_{\text{z}}}\left(\boldsymbol{r}_{\bot}, z, \omega\right) \right\}} \right)$. For a detailed description of the method, the setup and the various processing steps see supplement B and Refs.\,\cite{Nelayah2007, Mayer2019}.
	For analyzing the spatial localization of the LSP modes we corrected for elastic scattering absorption (i.e., scattering of electrons into large scattering angles) by normalizing the point-wise collected spectra with the overall intensity of each individual spectrum i.e., $\Gamma_{\text{n}}(\boldsymbol{r}_\perp, \omega)=\Gamma(\boldsymbol{r}_\perp, \omega)/\int \Gamma(\boldsymbol{r}_\perp, \omega)\,\text{d}\omega$ (see Fig.\,\ref{fig:spect_maps}\,a for a comparison of an as-recorded and an absorption corrected loss probability map).
	
	Examples of typical spectra, shown in Fig.\,\ref{fig:spect_maps}\,b, reveal the presence of spectrally localized SP modes of excitation energies $\hbar \omega$ ranging from approximately $0.3$\,eV (ca. $73$\,THz) to $1.7$\,eV (ca. $387$\,THz). Below $0.3$\,eV the limited energy resolution of the experiment prevented detection of modes. Most remarkably, the quality factor $Q=\omega/\Delta\omega$ given by the ratio of the energy of a particular mode and its energy width (FWHM) frequently reaches values in the range of $10$, which comes very close to the theoretical achievable maximum in Au limited by the comparatively large imaginary part (loss) of its bulk dielectric function (see supplement C). Displaying the spatial slices of the spectrum image data\-cube (Fig.\,\ref{fig:spect_maps}\,a,\,c) reveals the spatial localization of the SP modes (so-called hotspots), with the magnitude of the localization clearly increasing towards higher excitation energies. Moreover, we observe that the maximally localized modes (in the vicinity of the $1.7$\,eV threshold) are always centered around the underlying Au network stripes. In the region between approximately $1.7$\,eV and the surface plasmon cut-off frequency ($\omega_{\text{cut,\,Au}}$) of Au at $2.4$\,eV only very weak higher-order LSP modes of the Au-vacuum interface of almost absent mutual hybridization are observed (no hotspots). Accordingly, the mean loss probability $\overline{\Gamma}^{\text{exp}}_{\text{n}}(\omega)$ (averaged over several spatial subsets of the web) decreases until the threshold at ap\-proxi\-mately $1.7$\,eV and becomes very small in the spectral range between the threshold and $\omega_{\text{cut,\,Au}}$. The latter behavior is in good agreement with the optical transmission (see Fig.\,\ref{fig:corr_radial}\,a) which proves the coupling of the localized plasmons to free (transverse) electromagnetic waves. 

	\section{Anderson localization}
	
	By breaking any translational symmetry, the self-affine Au networks support localized modes, which are square integrable, and may be assigned a position and excitation energy. To measure the lo\-cali\-zation from the EEL spectrum images, we employed two different localization measures frequently used in the context of AL\,\cite{Zhu2020, Shi2018}: (I) the azimuthally averaged ($\langle...\rangle_\varphi$) autocorrelation
	\begin{align}
	R_{\text{exp}}\left(r,\omega\right)=\biggl\langle\int_{-\infty}^\infty&\left(\Gamma_{\text{n}}\left(\boldsymbol{r}_\perp+\boldsymbol{r}_\perp^{\prime},\omega\right)-\bar{\Gamma}_{\text{n}}(\omega)\right)\nonumber \\ &\left(\Gamma_{\text{n}}\left(\boldsymbol{r}_\perp^{\prime},\omega\right)-\bar{\Gamma}_{\text{n}}(\omega)\right)d^2r'_{\bot}\biggr\rangle_\varphi
	\end{align}
	 within a certain energy interval (midpoint $\omega$) corresponding to the energy resolution of the experiment, and (II) the inverse second momentum of the spectrum image
	 \begin{equation}
	  p_{\text{exp}}\left(\omega\right)=\left(\int_{-\infty}^\infty\frac{\Gamma_{\text{n}}^2\left(\boldsymbol{r}_\perp,\omega\right)}{\bar{\Gamma}^{2}_{\text{n}}(\omega)}d^2r_\bot\right)^{-1}
	 \end{equation}
	   within an energy interval (here, $\bar{\Gamma}_{\text{n}}(\omega)=\int\Gamma_{\text{n}}(\boldsymbol{r}_\perp, \omega) d^2r_\bot$ corresponds to the mean loss prob\-abil\-ity of a spatial subset at $\omega$). The latter is closely linked to the so-called participation number $p_{\text{sim}}$ and has been shown to reflect localization rather robustly independent of absorption and the character of localization (e.g., exponential or algebraic)\,\cite{Markel2006}.
	
\begin{figure}
		\centering
		\includegraphics{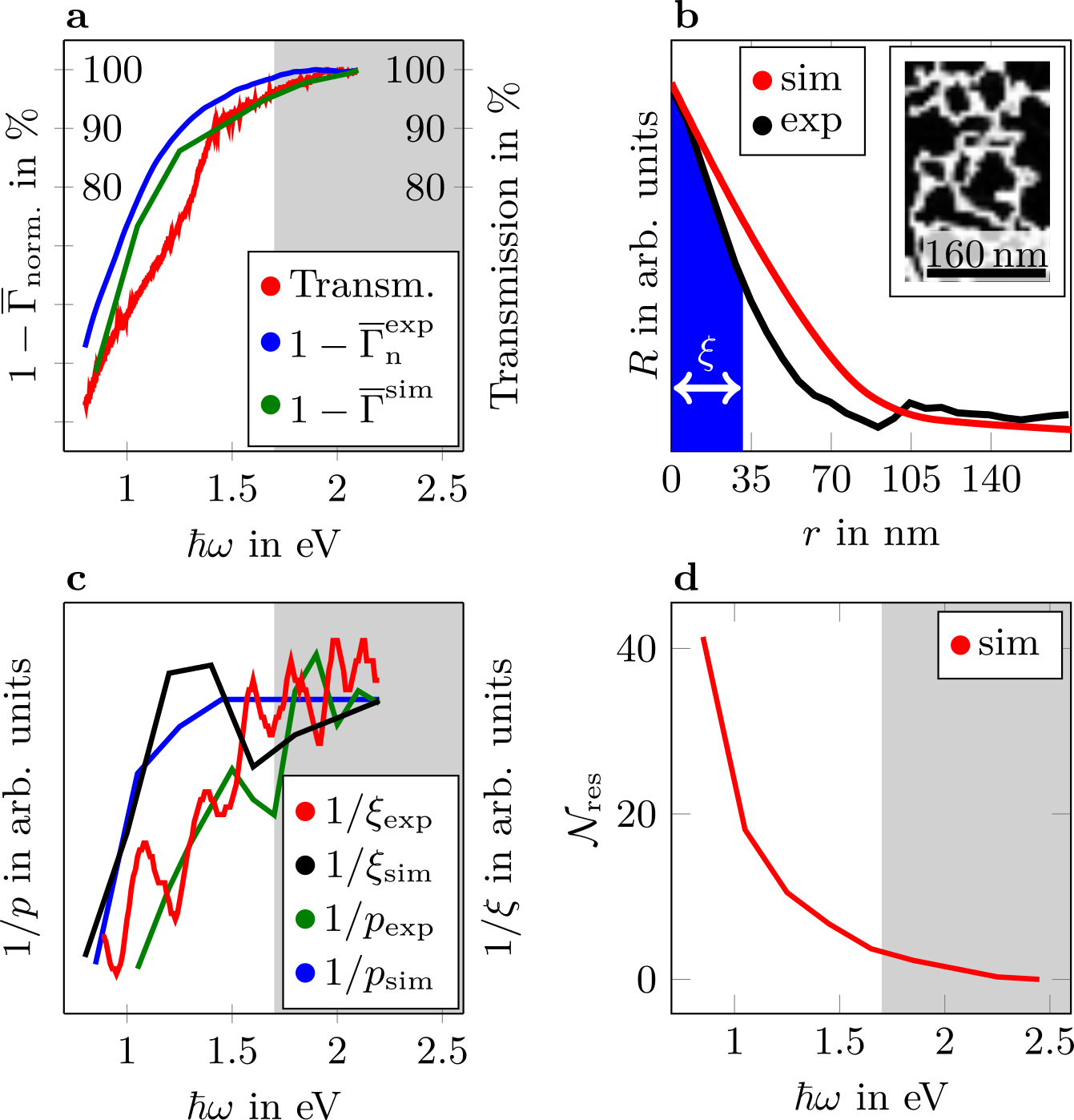}
		\caption{(a) Optical transmission of a macroscopic web (mm in size, coverage $\approx$\,0.4) compared with the spatially averaged loss probability $\overline{\Gamma}^{\text{exp}}_{\text{n}}(\omega)$ as well as the simulated loss probability $\overline{\Gamma}^{\text{sim}}(\omega)=\mathcal{N}_{\text{res}}(\omega)\left|\boldsymbol{\overline P}(\omega)\right|$. (b) Azimuthally averaged autocorrelation $R(r,\omega)$ of resonant LSP modes at 0.8\,eV energy loss. The width of the blue shaded area indicates the FWHM $\xi$ of the central peak (correlation length). (c) Spectral dependence of the inverse participation number $1/p(\omega)$ and correlation length $1/\xi(\omega)$. (d) Spectral dependence of the number of reso\-nant eigenmodes $\mathcal{N}_{\text{res}}(\omega)$ of the simulated system of coupled electric dipoles. The simulated data was averaged over an ensemble of 10 disorder configurations in all cases.}
		\label{fig:corr_radial}
	\end{figure}

	By exhibiting a characteristic central maximum and decreasing toward larger distances the autocorrelation confirms the localized nature of the surface plasmon reso\-nances (presence of hotspots). Their correlation length $\xi$ (as characterized by the Full Width at Half Maximum (FWHM, see Fig.\,\ref{fig:corr_radial}\,b) of the autocorrelation) decreases towards larger energies with no significant dependency on the network properties (see Fig.\,\ref{fig:corr_radial}\,c).
	The increasing localization of the LSPs is further corroborated by $p_{\text{exp}}$ (see Fig.\,\ref{fig:corr_radial}\,c), which increases toward higher energies until approximately $1.7$\,eV (i.e., the same energy where localized modes disappear (see Fig.\,\ref{fig:corr_radial}\,d).
	Please note that the evaluated spectral range is restricted to energies above $\approx0.8$\,eV due to the limited spatial extent of the meas\-ured spatial subsets eventually artificially cropping the modes.
	
	In order to further analyze the observed localization behavior, we conducted numerical simulations of surface plasmon resonances in which strength of disorder, loss, and retardation may be deliberately modified. The fully retarded response of the Au network is well described by macroscopic Maxwell's equations taken into account a frequency dependent inhomogeneous dielectric function, i.e.,  $\nabla\times\nabla\times\boldsymbol{E}\left(\boldsymbol{r},\omega\right)-k^{2}\varepsilon\left(\boldsymbol{r},\omega\right)\boldsymbol{E}\left(\boldsymbol{r},\omega\right)=-i\frac{4\pi k}{c}\boldsymbol{j}^\mathrm{ext}\left(\boldsymbol{r},\omega\right)$. Here $\boldsymbol{j}^\mathrm{ext}$ denotes the external current (e.g., electron beam in case of EELS), $k=\omega / c$ the wave number of the free photon, and we employed an experimentally determined complex dielectric function of Au $\varepsilon\left(\omega\right)=\varepsilon_\mathrm{Au}\left(\omega\right)$ in order to consider losses appropriately \cite{diel_Au}.

	The prefactor $k^2$ of the spatially random dielectric function increases the impact of disorder at larger $\omega$ (additional frequency dependency is introduced by the dielectric function), which agrees well with the experimentally observed increase of localization. This behavior is markedly different from quantum AL, where such an "intrinsic" amplification of disorder toward higher frequencies is absent and disorder is dominant in the low energy limit \cite{Abrahams2010}. Another fundamental difference to AL within the framework of Schr\"odinger equation is vectorial character of the electric field \cite{Skipetrov2014}.
	
	In order to circumvent the computationally demanding (if not unfeasible) computation of the above partial differential equation on a large Au net, we solved the geo\-met\-ri\-cal\-ly inverse problem of interacting randomly distributed oblate Au nanoellipsoids (representing the holes of the network) of random lateral main axis and same thickness as the network yielding the same (inverse) coverage like the Au network. This approach exploits the qualitative similarity between the plasmonic response of the complementary and the original system due to a generalized Babinet principle\,\cite{Horak2019, Zentgraf2007, Falcone2004}. The latter holds in the limit of thin samples (in comparison to the photon wavelength) consisting of perfectly conducting material\,\cite{Horak2019}. Both conditions are satisfied well by the investigated 2D networks in the evaluated frequency range ($\omega$ well below the material's plasma frequency $\omega_p$). Restricting the interaction between the NPs to the dominant dipole coupling leads to further simplification, i.e., the following discrete equation system for the particles' dipole moments $\boldsymbol{P}$:
	\begin{equation} \boldsymbol{P}_{i}\left(\omega\right)-\boldsymbol{\alpha}_{i}\left(\omega\right)\sum_{j=1,\,j\neq i}^{N}\mathbf{G}_{ij}\left(\omega\right)\boldsymbol{P}_{j}\left(\omega\right)=\boldsymbol{\alpha}_{i}\left(\omega\right)\boldsymbol{E}^{\mathrm{ext}}_{i}\left(\omega\right) \nonumber
	\end{equation}	
	 with the retarded dipole interaction \cite{novotny_hecht_2012}
	\begin{align} \mathbf{G}_{ij}\left(\omega\right)&=\frac{e^{ikr_{ij}}k^{2}}{4\pi\varepsilon_{0}r_{ij}} \cdot  \left(\mathbf{I}_3-\boldsymbol{e}_{ij}\otimes \boldsymbol{e}_{ij}\right) \nonumber \\
	&+\frac{e^{ikr_{ij}}\left(1-ikr_{ij}\right)}{4\pi\varepsilon_{0}}\left(\frac{3\boldsymbol{e}_{ij}\otimes\boldsymbol{e}_{ij}-\mathbf{I}_3}{r_{ij}^{3}}\right).\nonumber
	\end{align}
	\begin{figure*}
		\centering
		\includegraphics{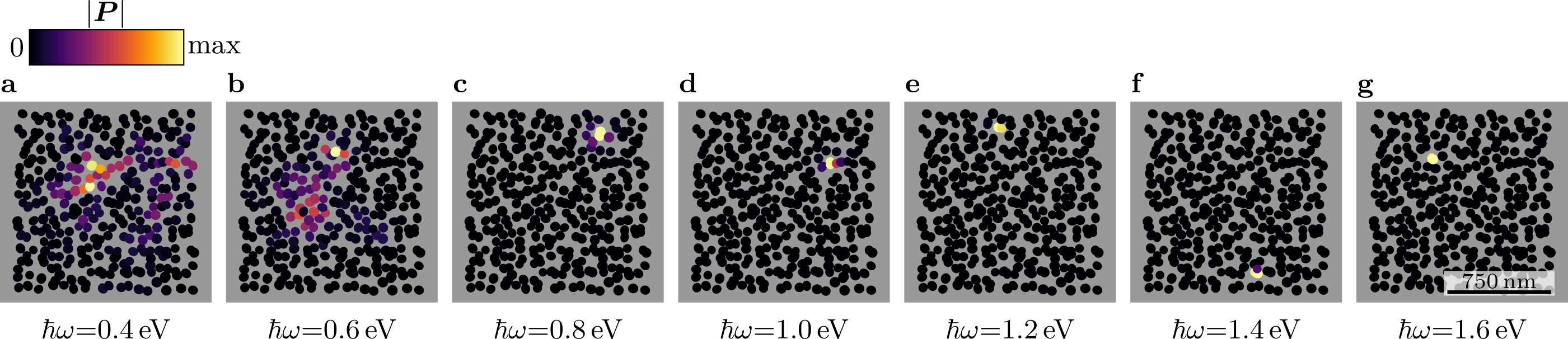}
		\caption{Spatial distribution of the induced dipole moments of selected resonant eigenmodes of the simulated system of coupled dipoles at different energies. The number of dipoles participating to a resonant mode (participation number) decreases with increasing energy revealing stronger localization with higher energy.}
		\label{fig:sim_maps}
	\end{figure*}
	Here the indices $i,j$, running from $1$ to the number of particles $N$, denote the individual NPs, $\mathbf{I}_3$ the $3\times 3$ identity matrix, $\boldsymbol{\alpha}_i$ the polarizability tensor of an ellipsoidal NP, $\boldsymbol{E}^{\mathrm{ext}}_{i}$ the external electric field (e.g., corresponding to the Li\'enard-Wiechert potential of the beam electrons), and $\boldsymbol{e}_{ij}$ the unit distance vector between two NPs (see supplement D for details).
	
	Resonant modes occur when  $\left(\mathbf{I}_{3N}-\boldsymbol{\alpha}\mathbf{G}\right)$ approaches zero (i.e., a small external field leads to a large response), where we wrapped all the $\boldsymbol{\alpha}_i$ into one matrix $\boldsymbol{\alpha}$ of size $3N$ and abbreviated the application of the dipole interaction over all particle indices into one large $\mathbf{G}$. In the following, we will determine these modes by solving the corresponding eigenvalue problem $\mathcal{P}\left(\omega\right)\boldsymbol{P}_\mathcal{P}\left(\omega\right)=\boldsymbol{\alpha}\left(\omega\right)\mathbf{G}\left(\omega\right)\boldsymbol{P}_\mathcal{P}\left(\omega\right)$ in dependence of $\omega$ and restricting the eigenspace to modes with complex eigen\-values $\mathcal{P}$ within a small interval around 1 (resonance condition, see supplement D), i.e., $\left|\mathcal{P}-1\right| \leq \delta$. Their lo\-cali\-za\-tion behavior is then analyzed by computing their azimuthally averaged $\left\langle ... \right\rangle_{\varphi}$ autocorrelation
	\begin{align}
	 R_{\text{sim}}\left(r_\perp,\omega\right)=\biggl\langle\int&\left(\boldsymbol{P}_\mathcal{P}\left(\boldsymbol{r}_\perp+\boldsymbol{r}_\perp^{\prime},\omega\right)-\bar{\boldsymbol{P}}_\mathcal{P}(\omega)\right)\nonumber\\ &\left(\boldsymbol{P}_\mathcal{P}\left(\boldsymbol{r}_\perp^{\prime},\omega\right)-\bar{\boldsymbol{P}}_\mathcal{P}(\omega)\right)d^2r'_{\bot}\biggr\rangle_{\mathcal{P}, \varphi}
	 \end{align}
	and participation number $p_{\text{sim}}(\omega)=\left\langle \sum_{i=1}^{N}\frac{1}{\left|\boldsymbol{P}_{i, \mathcal{P}}\left(\omega\right)\right|^{2}}\right\rangle _{\mathcal{P}}$ from the induced dipole moments of the eigenmodes $\boldsymbol{P}_\mathcal{P}$ and the corresponding mean value $\bar{\boldsymbol{P}}_\mathcal{P}(\omega)=\int\boldsymbol{P}_\mathcal{P}(\boldsymbol{r}_\perp, \omega)d^2r_{\bot}$. Both, $R_{\text{sim}}$ and $p_{\text{sim}}$ are averaged over resonant modes in the interval around 1 indicated by $\langle...\rangle_\mathcal{P}$, respectively.
	
	Accordingly, the eigensolutions in close vicinity to the resonance condition are localized depending on the corresponding energy (see Fig.\,\ref{fig:sim_maps}\,a-g) in good agreement with the experiment (Fig.\,\ref{fig:spect_maps}\,c). Similarly, the simulated inverse participation number increases with energy as observed experimentally (Fig.\,\ref{fig:corr_radial}\,c). Moreover, the number of resonant eigenmodes decrease toward higher energies and is eventually completely suppressed above a spectral threshold around $1.7$\,eV-$1.8$\,eV (see Fig.\,\ref{fig:corr_radial}\,d). The latter reflects the experimental observation of vanishing hybridized LSP modes above this threshold (see Fig.\,\ref{fig:spect_maps}\,b,c). To calculate a measure for the loss probability we multiplied the number of resonant modes $\mathcal{N}_{\text{res}}(\omega)$, corresponding to the optical density of states, with the mean induced dipole moment $\left|\boldsymbol{\overline P}(\omega)\right|$ (averaged over all coupled dipoles and resonant modes), corresponding to the inter\-ac\-tion strength ($\overline \Gamma^{\text{sim}}(\omega)=\mathcal{N}_{\text{res}}(\omega)\left|\boldsymbol{\overline P}(\omega)\right|$). The obtained quantity reproduces the experimental loss prob\-abil\-ities very well (see Fig.\,\ref{fig:corr_radial}\,a). Below $1.7$\,eV $\Gamma^{\text{sim}}$ decreases with higher energy and saturates at small values above the $1.7$\,eV threshold reflecting again the high optical transparency in the visible spectral range. Separate analysis of the different vector components of the induced dipole moments $\boldsymbol{P}$ identified the in-plane components as the dominant driving force of the observed localization (see Fig.\,S2).

	Having established a good qualitative and partly even quantitative agreement between experimentally observed LSPs in the Au networks and the simulated LSPs in the randomly distributed oblate NPs, we may now draw the following conclusions about the localization behavior:\\ (I) Switching off loss and retardation (see Fig.\,S3 in supplement E) does not significantly alter the localization behavior. While the former rules out finite lifetime related localization effects as dominant mechanism behind the observed localization the latter shows that wave interference on short distances below the photon wavelength is the driving force (in correspondence to Ioffe-Regel criterion). The last observation is corroborated by the fact that the ordered square lattice of identical NPs (corresponding to a regular network of holes) sustains quasi-continuous plasmon bands (not shown)\,\cite{Mayer2019}.\\
	(II) The localization behaviour in this 2D system is driven by the positional disorder of the holes and not their size disorder (within boundaries given by simulated coverage rates, see Fig.\,S4 in supplement E). The latter property may significantly change upon increasing the thickness, also modifying the localization behaviour as a consequence.\\
	(III) The disappearance of hybridized localized modes above a spectral threshold ($1.7$\,eV in case of the investigated Au webs) independent of loss magnitude is a universal localization effect in the following sense. Different geometric parameters of the network (i.e., coverage, size of holes) or material composition may lead to a shift of the spectral threshold whereas the general lo\-cali\-za\-tion behavior remains (see Fig.\,S3 in supplement E). For instance, Au NP assemblies of lower coverage exhibit a lower threshold, corresponding to smaller/less holes in the network (see blue curve in Fig.\,S3 in the supplement). \\
	We therefore attribute the suppression of LSPs in these networks to a (destructive) wave interference effect ultimately canceling the dominant dipolar coupling between various hot spots above a certain frequency. This disappearance of localized modes above the threshold at the lower end of the optical spectrum explains the exceptionally large transparencies up to 97\,\% (in the optical frequency range 443\,THz\,-\,635\,THz) of the Au networks\,\cite{Hiekel2020}. We note, however, that our data does not allow a statement about a possible existence of a mobility edge (i.e. the transition to fully delocalized states) below $0.3$\,eV. Reasons are the limited energy resolution of the experiment and the system size limits of the simulations.
\section{Summary and Outlook}

Summing up, we showed both experimentally and theo\-reti\-cal\-ly that (self-affine) metallic networks of very low mass thickness and coverage support Anderson localized (disorder-driven) LSP resonances that are confined to the network plane in their oscillation direction. These LSP modes have very large quality factors close to the theoretical maximum and show increasing spatial localization toward higher excitation energies (frequencies). They typi\-cal\-ly consist of few hotspots, couple to electromagnetic plane waves, and their localization behaviour is driven by the random distribution of the network holes mainly. They are only weakly affected by retardation effects and disappear above ca. 1.7\,eV in the investigated Au networks, which explains their exceptionally large transparency in the optical spectrum.\\
Unresolved questions pertain to the existence of a mobility edge and generally to a better understanding for the observed correlation distance, their spectral distribution, and the impact of the networks' self-affinity and thickness, which hinges on the development of an analytical description of AL in such random 2D networks (e.g., via self-consistent AL theory). Such a theory could support the development of design rules for the networks, e.g., in order to further optimize design transparency for envisaged future application as transparent electrodes.

    \bibliographystyle{apsrev4-2}
	\bibliography{plasmon_loc}

	\begin{acknowledgments}
		 J.S. received funding from the Deutsche Forschungsgemeinschaft (DFG, German Research Foundation)-project-id 461150024. K.H. received funding from the European Union’s Proof of Concept project LAACat no. 875564, from the European Research Council through the ERC AdG 2013 AEROCAT, and the Graduate Academy of TU Dresden. P.P. received funding from the Deutsche Forschungsgemeinschaft (DFG, German Research Foundation)-project-id 431448015. A.L. received funding from the Deutsche Forschungsgemeinschaft (DFG, German Research Foundation)-project-id 417590517. P.K. and A.E. are grateful to the DFG for support through DFG EY 16/30-1 and RTG 2767. The data accompanying this publication are available in \cite{Schultz2023}.
	\end{acknowledgments}

\newpage
\begin{appendix}
	\section{Synthesis of the Au network}
	The general synthesis route of the two-dimensional Au networks (see Fig.\,\ref{fig:overview} for an example) corresponds to that described in Hiekel at al.\cite{Hiekel2020} First, Au NPs were synthesized. To that end, 0.1\,mmol HAuCl$_4\cdot$3H$_2$O (Sigma-Aldrich\,>\,99.9\,\% trace metal basis) was dissolved in 492\,ml water and solutions (volumes 2 till 4\,ml) of 0.143\,mmol/ml NaBH$_4$ (Sigma-Aldrich,\,>\,96\,\%) were swiftly added before stirring the solution for 30\,min. In one case, the obtained solution was diluted afterwards by adding 100 ml water.
	
	Subsequently, the two-dimensional Au network structures were prepared in the following way. Firstly, 200\,$\mu$l or 400\,$\mu$l of the above synthesized Au NP solution were deposited on cover slips (24 x 24\,mm, washed with ac\-etone). Subsequently, the solution was slowly overlaid with 100\,$\mu$l of a toluene/EtOH (1:1) mixture and the two-dimensional Au network structures formed at the phase boundary. After the evaporation of the organic solvent the structures were transferred to a TEM-grid or another cover slip by carefully pressing the substrate onto the aqueous solution. Then, the substrate was washed with EtOH. Optical transmission spectra were recorded using a Varian Cary 5000 absorption spectrometer.
	\renewcommand{\thefigure}{S1}
	\begin{figure}[h]
		\centering
		\includegraphics{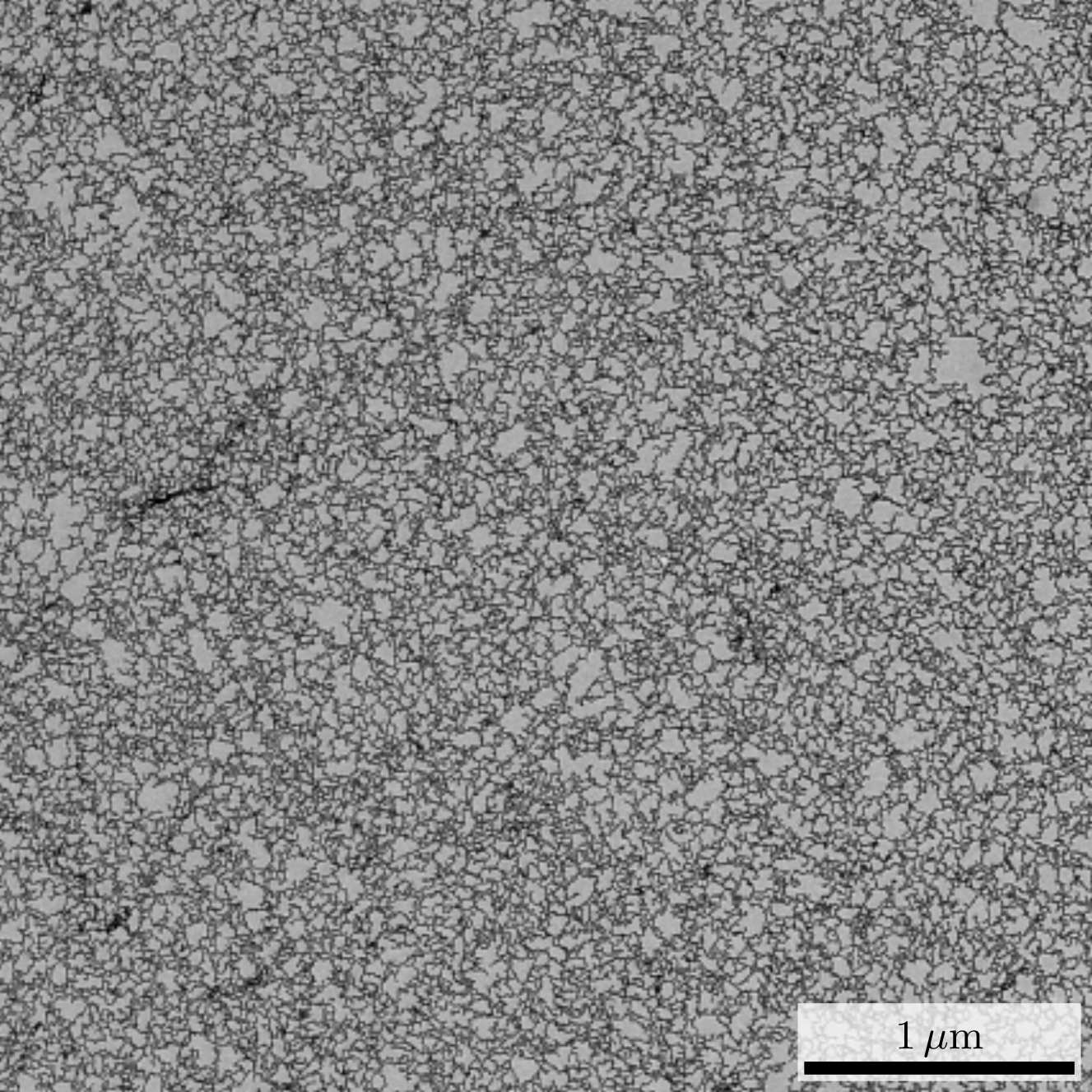}
		\caption{Transmission Electron Microscopy Image of an exemplary two-dimensional Au network.}
		\label{fig:overview}
	\end{figure}
	
	\section{Plasmon mapping with STEM-EELS}
	STEM scanning combined with acquisition of EELS spectra was performed in the probe-corrected FEI Titan$^{3}$ TEM operating at 80 kV acceleration voltage. The micro\-scope is equipped with a Gatan Tridiem energy filter and a Wien monochromator, which was optimized\,\cite{Lopatin2018} to fa\-cili\-tate an energy resolution of 60-70 meV. EELS was performed under a convergence angle of 16 mrad and a collection angle of 3 mrad with an energy dispersion of 0.01 eV per channel. The spectrum images were acquired with a beam current of 120 pA and dwell time of 30 ms. To improve measuring statistics, several (up to 10) repeated spectrum-images were recorded for each dataset followed by the summation of the data  accounting for individual spatial drifts. The obtained spectra were corrected for the temporal energy instability of primary electrons through the alignment of the zero-loss peak (ZLP).  This peak, prevailing in the low-loss EELS region, was then deconvolved from the spectra by means of Richardson-Lucy algorithm and ultimately subtracted employing a reference profile collected in a separate run in vacuum. Finally, the energy positions and magnitude of distinct peaks in the low-loss region were fitted using the nonlinear least-squares procedure to determine the quality factors.
	
	\section{Theoretical Limit of the $Q$-factor}
	In general the $Q$-factor of a LSP mode is defined as $Q=\frac{\omega}{\Delta\omega}$ where $\hbar\Delta\omega$ corresponds to the full with at half maximum (FWHM) of the spectral peak in the loss spectrum. From the time domain Fourier transform of the damped plasmonic oscillator (with damping constant $\tau$) one obtains a Lorentzian line shape: $\mathfrak{F}\left\{ e^{-a t} \right\}=\mathfrak{F}\left\{ e^{-i\omega t} \cdot e^{-t/\tau}\right\}= \sqrt{\frac{1}{2\pi}}\frac{1/\tau}{\left(1/\tau\right)^{2}+(\omega-\omega_0)^2}$. Ana\-lyz\-ing the extrema of the Lorentzian curve the maximum can be found at $\omega-\omega_0=0$. Hence, it follows for the FWHM $\hbar\Delta\omega=\frac{\hbar}{2\tau}$. Assuming $\tau=5.646\cdot10^{-14}$\,s\,\cite{Berciaud} this directly leads to $\hbar\Delta\omega\approx20$\,meV. At $1$\,eV loss energy this corresponds to $Q\approx45$ which represents the theoretical limit which is reduced in practice due to loss channels not considered in the Drude model e.g., radiative losses, inter- and intraband transitions, etc. We note furthermore that the experimentally determined FWHM is also broadened due to the residual width of the zero loss peak in the EEL spectrum of $\approx30$\,meV after the Richardson-Lucy deconvolution. Consequently, the experimentally observed quality factors of the AL localized LSPs are close to the theoretical maximum in Au plasmonic nanostructures, which may be further verified by comparing to reported $Q$-factors in the literature (e.g., \cite{Horak2019}).
		
	\section{Simulation Details}
	
	The self-consistent dipole coupling model
	
	\begin{equation}
		\boldsymbol{P}_{i}\left(\omega\right)=\boldsymbol{\alpha}_{i}\left(\omega\right)\left(\boldsymbol{E}_{\mathrm{ext},i}\left(\omega\right)-\sum_{j=1,\,j\neq i}^{N}\mathbf{G}_{ij}\left(\omega\right)\boldsymbol{P}_{j}\left(\omega\right)\right)\label{eq:SDCM}
	\end{equation}
	couples a set of $N$ NPs, denoted by $i$ or $j$, with anisotropic
	polarizability tensor $\boldsymbol{\alpha}_{i}$ via dipole interaction \cite{novotny_hecht_2012}
	\begin{align}
		\mathbf{G}_{ij}\left(\omega\right)=&\frac{e^{ikr_{ij}}}{4\pi\varepsilon_{0}}\frac{k^{2}}{r_{ij}}\left(\mathbf{I}_{3}-\boldsymbol{e}_{ij}\otimes\boldsymbol{e}_{ij}\right)+\\
		&\frac{e^{ikr_{ij}}}{4\pi\varepsilon_{0}}\frac{\left(1-ikr_{ij}\right)\left(3\boldsymbol{e}_{ij}\otimes\boldsymbol{e}_{ij}-\mathbf{I}_{3}\right)}{r_{ij}^{3}}\nonumber \label{Green}
	\end{align}
	and an external electric field $\boldsymbol{E}_{\mathrm{ext},i}$
	(in our case the evanescent field produced by the scanning electron beam). Here, the	wave number reads $k=\omega/c$ and the interparticle unit distance
	vector is $\boldsymbol{e}_{ij}=\boldsymbol{r}_{ij}/r_{ij}$. 
	
	In order to test the impact of retardation on the localization behaviour we also tested the non-retarded version of the dipole interaction \cite{novotny_hecht_2012}
	
	\begin{equation}
		\mathbf{G}_{ij}=\frac{1}{4\pi\varepsilon_{0}}\frac{3\boldsymbol{e}_{ij}\otimes\boldsymbol{e}_{ij}-\mathbf{I}_{3}}{r_{ij}^{3}}\,.
	\end{equation}
	As we seek to model the geometry of the holes in the two-dimensional random networks, we approximate the NPs as highly oblate ellipsoids, for which an analytic
	solution for the polarizability tensor exists. Along the principle
	axis of an ellipsoid (denoted by $m$) it reads \cite{bohren1983}
	\begin{equation}
		\alpha_{mm}\left(\omega\right)=4\pi\varepsilon_{0}a_{1}a_{2}a_{3}\frac{\varepsilon\left(\omega\right)-1}{3+3L_{i}\left(\varepsilon\left(\omega\right)-1\right)}
	\end{equation}
	with the purely geometrical depolarization factors
	\begin{equation}
		L_{m}=\frac{a_{1}a_{2}a_{3}}{2}\int_{0}^{\infty}\frac{1}{\left(a_{m}^{2}+q\right)\sqrt{\prod_{n=1}^{3}\left(q+a_{n}^{2}\right)}}dq
	\end{equation}
	satisfying $L_1+L_2+L_3=1$. Here, $a_{m}$ are the semiaxes of the ellipsoid and $\varepsilon\left(\omega\right)$
	the bulk dielectric function of the metal. In our simulations, we
	will always employ strongly oblate ellipsoids with in-plane semiaxes $a_{1},\,a_{2}$ in the range of several tens of nanometers and out-of-plane $a_{3}=a_z$ below 1 nm, i.e., $a_{1},\,a_{2}\gg a_{3}$ by roughly one order of magnitude, which corresponds well with the observed nature of the holes. To include radiative corrections we also incorporated the first order correction factor according to \cite{novotny_hecht_2012}, i.e., 	
		\begin{equation}
			\boldsymbol{\alpha}_{i} \to \frac{\boldsymbol{\alpha}_{i}}{1-\frac{k^3}{6\pi\epsilon_0}\boldsymbol{\alpha}_{i}}\,.
		\end{equation}
		As a consequence of $k^3 \prod_m a_m \ll 1$, however, this correction is small and may be also neglected, which we explicitely checked in the simulations by switching retardation on and off (see Fig. \ref{fig:6}). In order to compute the polarizability tensor for arbitrary in-plane orientations of the oblate ellipsoids (i.e., with in-plane principle axis not coinciding with the coordinate axis), the diagonal $\boldsymbol{\alpha}$ matrix is rotated by left and right multiplication with a $xy$ rotation matrix. 
	
	The dipole coupling model (\ref{eq:SDCM}) may be written in the following compact form
	\begin{equation}
		\left(\boldsymbol{\alpha}^{-1}\left(\omega\right)+\mathbf{G}\left(\omega\right)\right)\boldsymbol{P}\left(\omega\right)=\boldsymbol{E}_{\mathrm{ext}}\left(\omega\right)
	\end{equation}
	where $\boldsymbol{\alpha}$ is the $3N\times3N$ matrix of
	all NP polarizability tensors
	\begin{equation}
		\boldsymbol{\alpha}=\begin{pmatrix}\boldsymbol{\alpha}_{1} & \boldsymbol{0} & \cdots & \boldsymbol{0}\\
			\boldsymbol{0} & \boldsymbol{\alpha}_{2} & \ddots & \boldsymbol{0}\\
			\vdots & \ddots & \ddots & \vdots\\
			\boldsymbol{0} & \boldsymbol{0} & \cdots & \boldsymbol{\alpha}_{N}
		\end{pmatrix}
	\end{equation}
	and $\mathbf{G}$ the matrix operator of all dipole interactions
	\begin{equation}
		\mathbf{G}=\begin{pmatrix}\boldsymbol{0} & \mathbf{G}_{12} & \cdots & \mathbf{G}_{1N}\\
			\mathbf{G}_{21} & \boldsymbol{0} & \ddots & \mathbf{G}_{2N}\\
			\vdots & \ddots & \ddots & \vdots\\
			\mathbf{G}_{N1} & \mathbf{G}_{N2} & \cdots & \boldsymbol{0}
		\end{pmatrix}\,.
	\end{equation}
	The above model is a simplified version of the more general MESME
	model, which employs a higher order multi\-pole expansion beyond dipolar coupling \cite{GarciadeAbajo1999}. We show in the following that the pure dipolar coupling model indeed reproduces the main observations, while being sufficiently simple to facilitate analytical transformations as well as numerical solutions. Notwithstanding the impact of higher-order coupling may lead to additional spectral shifts as well as localization effects, not considered here.
	
	The singularities of the resolvent $\mathbf{R}\left(\omega\right)=\left(\boldsymbol{\alpha}^{-1}\left(\omega\right)-\mathbf{G}\left(\omega\right)\right)^{-1}$,
		define the frequencies at which the dipole assembly response is resonant. These frequencies are complex in general, whereas real  frequencies or energies are probed in the EELS experiment (we will come back to that point further below). The resonances can be found by searching for zeros of $\mathbf{I}_{3N}-\boldsymbol{\alpha}\left(\omega\right)\mathbf{G}\left(\omega\right)$
		as a function of $\omega$.  
		To circumvent the costly zero search including
		the computation of associated modes, we assume that the matrix $\mathbf{A}\left(\omega\right):=\boldsymbol{\alpha}\left(\omega\right)\mathbf{G}\left(\omega\right)$ is not defective and thus has a complete set of eigenvectors
		\begin{equation}
			\boldsymbol{\alpha}\left(\omega\right)\mathbf{G}\left(\omega\right)\boldsymbol{P}_k\left(\omega\right)=\mathcal{P}_k\left(\omega\right)\boldsymbol{P}_k\left(\omega\right)\,,
		\end{equation}
		that guarantees the existence of a dual basis $\left\langle \boldsymbol{P}_k\left(\omega\right)\vert\boldsymbol{Y}_l\left(\omega\right)\right\rangle=\boldsymbol{Z}_k\left(\omega\right)\delta_{kl}$ at each frequency $\omega$. The resolvent can then be written as
		\begin{equation}
			\mathbf{R}\left(\omega\right)=\sum_k\frac{\left\vert\boldsymbol{P}_k\left(\omega\right)\right\rangle\left\langle\boldsymbol{Y}_k\left(\omega\right)\right\vert}{\boldsymbol{Z}_k\left(\omega\right)\left(1-\mathcal{P}_k\left(\omega\right)\right)}.
		\end{equation}	
		Therefore the singularities of the resolvent occur at frequencies where $\mathcal{P}_k\approx1$. Hence, we can consider those modes resonant, which have eigenvalues that fall into a small interval around 1, i.e., $\left|\mathcal{P}_k-1\right| \leq \delta$.
	In our analysis we deliberately set $\delta=0.4$ in order to have
	sufficiently large statistics for the autocorrelation and inverse
	participation number, when averaging over the ensemble of resonant
	modes. Smaller intervals yield similar results in terms of localization
	with larger stochastic noise. Larger intervals should be avoided in order
	to not pick up non-resonant modes close to the accumulation point
	of the spectrum at 0.
	
	Using that model we computed the localization properties of the resonant
	modes (typically as an average over all resonant modes in the interval) for a set of free pa\-ra\-me\-ters of the model. In order to improve the statistics, we generally averaged the results over several lattice configurations. As reference structure we employed
	a square lattice of NPs (i.e., holes) matching the filling factor
	of the experiment. We then varied
	\begin{itemize}
		\item geometrical parameters:
		\begin{itemize}
			\item total size of the system (0.75$\mu$m\,-\,1.5$\mu$m)
			\item filling factor/coverage via density of NPs (18$^2$-29$^2$ NPs (representing the holes of the webs) in the system leading to filling factors of 0.1\,-\,0.4)
			\item disorder via uniform principal axis ($\pm$ 10\,nm\,-\,20\,nm) and orientation distribution of the NPs (corresponding to random diagonal entries in $\boldsymbol{\alpha}$) and uniform position distribution (corresponding to off-diagonal random entries of $\mathbf{G}$)
		\end{itemize}
		Please note that by independently randomizing the principal axis length of the NPs also their volume was randomized. The uniform distribution of the NPs around their nominal positions of the square lattice was limited such that the square distribution boxes of adjacent NPs touched but did not overlap.
		\item dielectric parameters ($\varepsilon_{\text{gold}}$ and $\varepsilon_{\text{aluminum}}$ are taken from \cite{diel_Au} and \cite{diel_Al})
		\item interaction parameters (quasi-static, fully retarded)
	\end{itemize}
	and evaluated the autocorrelation and inverse participation number
	(and other characteristics): For the autocorrelation,
	we evaluated 
	\begin{align}
		R_{\text{sim}}(r_\perp,\omega)=\biggl\langle\int&\left(\boldsymbol{P}_\mathcal{P}\left(\boldsymbol{r}_\perp+\boldsymbol{r}_\perp^{\prime},\omega\right)-\bar{\boldsymbol{P}}_\mathcal{P}(\omega)\right)\nonumber\\ &\left(\boldsymbol{P}_\mathcal{P}\left(\boldsymbol{r}_\perp^{\prime},\omega\right)-\bar{\boldsymbol{P}}_\mathcal{P}(\omega)\right)d^2r'_{\bot}\biggr\rangle_{\mathcal{P}, \varphi}
	\end{align}
	as a function of $\omega$ and $\boldsymbol{r}_\perp$ (averaged over the azimuthal coordinate $\varphi$ as well as resonant eigenmodes $\mathcal{P}$ in the interval around 1). Here, $\boldsymbol{P}_\mathcal{P}$ denotes the electrical fields of the eigenmodes with the corresponding mean value $\bar{\boldsymbol{P}}_\mathcal{P}(\omega)=\int\boldsymbol{P}_\mathcal{P}(\boldsymbol{r}_\perp, \omega)d^2r_{\bot}$. The participation number was obtained
	through
	\begin{equation}
		p_{\text{sim}}(\omega)=\left\langle \sum_{i=1}^{N}\frac{1}{\left|\boldsymbol{P}_{i, \mathcal{P}}\left(\omega\right)\right|^{2}}\right\rangle _{\mathcal{P}}
	\end{equation}
	again averaged over all resonant eigenmodes.
	The whole algorithm has been implemented in Julia programming language employing efficient libraries for linear algebra and numerical integration.

	\section{Impact of Geometry, Damping and Retardation on Localization}
	
	We first note that the localization behavior is strongly dominated by the in-plane ($\bot$) field components, since the mean dipole moment ($\overline{P}$, averaged over all nanoplatelets and resonant eigenmodes) perpendicular to the nanooblates is negligible in com\-pari\-son to the in-plane components (see Fig.\,\ref{fig:7}).
	\renewcommand{\thefigure}{S2}
	\begin{figure}[h]
		\centering
		\includegraphics{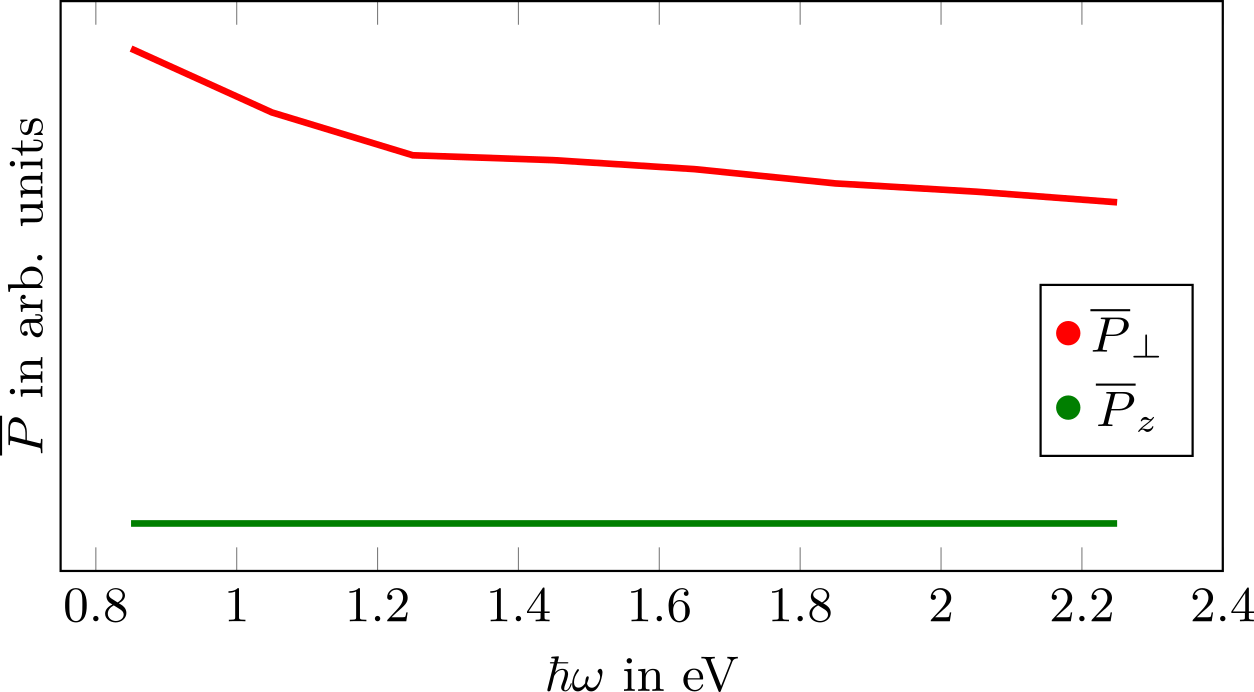}
		\caption{Comparison of the mean dipole moment in-plane ($\overline{P}_{\bot}$) and perpendicular to the nanooblates ($\overline{P}_z$).}
		\label{fig:7}
	\end{figure}
	
	This observation in combination with $L_1 \approx L_2 \approx 0 \ll L_3 \approx 1$ following from $a_{1},\,a_{2}\gg a_{3}$ motivates the  following separation of in-plane and out-of-plane components. We start out by rephrasing the inverse polarizability
	
	\begin{align}
		\boldsymbol{\alpha}^{-1}_i\left(\omega\right)&=\frac{1}{V_i}\chi\left(\omega\right) \nonumber \\ &+\frac{1}{V_i}\sum_{n=1}^3\left(3L_{i,n}-1\right)\boldsymbol{e}_{i,n}\otimes\boldsymbol{e}_{i,n} \nonumber \\ 
		&-i\frac{3k^3}{3}\mathbf{I}_3\, ,
	\end{align}
	where $V_i=a_{i1}a_{i2}a_{i3}$ is the volume of the NPs, $\boldsymbol{e}_{i,n}$ the unit vector along principal axis of $i$th NP, and
	
	\begin{equation}
		\chi\left(\omega\right) :=  \frac{\epsilon\left(\omega\right)+2}{\epsilon\left(\omega\right)-1}\,.
	\end{equation}
	Substituting into Eq. (\ref{eq:SDCM}), while taking into account $L_1 \approx L_2 \approx 0 \ll L_3 \approx 1$ and $G_{ij,13}=G_{ij,23}=0$, decouples the resulting equation in the $xy$- and $z$-plane. The $xy$-plane equation, hosting the localization behaviour as noted above, reads
	
	\begin{equation}
		\left(\chi\left(\omega\right)-1\right)\boldsymbol{P}_i=V_i\boldsymbol{E}_i+V_i\left(\mathbf{G}_{ij,\perp}\left(\omega\right)+\frac{2k^3}{3}\mathbf{I}_2\right)\boldsymbol{P}_j\label{eq:SDCM2D}
	\end{equation}
	with
	\begin{equation}
		\mathbf{G}_{ij,\perp} = \left(\begin{matrix}
			G_{ij,11}& G_{ij,12} \\
			G_{ij,21}& G_{ij,22}
		\end{matrix}\right)  \, .	
	\end{equation}	
	Using a Drude dielectric function
	
	\begin{equation}
		\chi\left(\omega\right)-1=-3\frac{\omega\left(\omega+i/\tau\right)}{\omega_p^2}
	\end{equation}
	the resonance condition reads
	
	\begin{equation}
		\left|3\frac{\omega\left(\omega+i/\tau\right)}{\omega_p^2}+\lambda_k\left(\omega\right)\right|\leq\delta\, ,
	\end{equation}
	where $\lambda_k\left(\omega\right)$ are eigenvalues of the matrix on the right hand side of Eq. (\ref{eq:SDCM2D}) defined by 
	\begin{equation}
		\mathbf{D}\left(\mathbf{G}_\perp\left(\omega\right)+\frac{2k^3}{3}\mathbf{I}_{2N}\right)\boldsymbol{P}_k=\lambda_k \boldsymbol{P}_k \label{eq:eigSDCM2D}
	\end{equation}
	with 
	\begin{equation}
		\mathbf{D} := \begin{pmatrix}V_1\mathbf{I}_{2} & \boldsymbol{0} & \cdots & \boldsymbol{0}\\
			\boldsymbol{0} & V_2\mathbf{I}_{2} & \ddots & \boldsymbol{0}\\
			\vdots & \ddots & \ddots & \vdots\\
			\boldsymbol{0} & \boldsymbol{0} & \cdots & V_N\mathbf{I}_{2}
		\end{pmatrix}\, .
	\end{equation}
		\renewcommand{\thefigure}{S3}
	\begin{figure}[!h]
		\centering
		\includegraphics[width=0.42\textwidth]{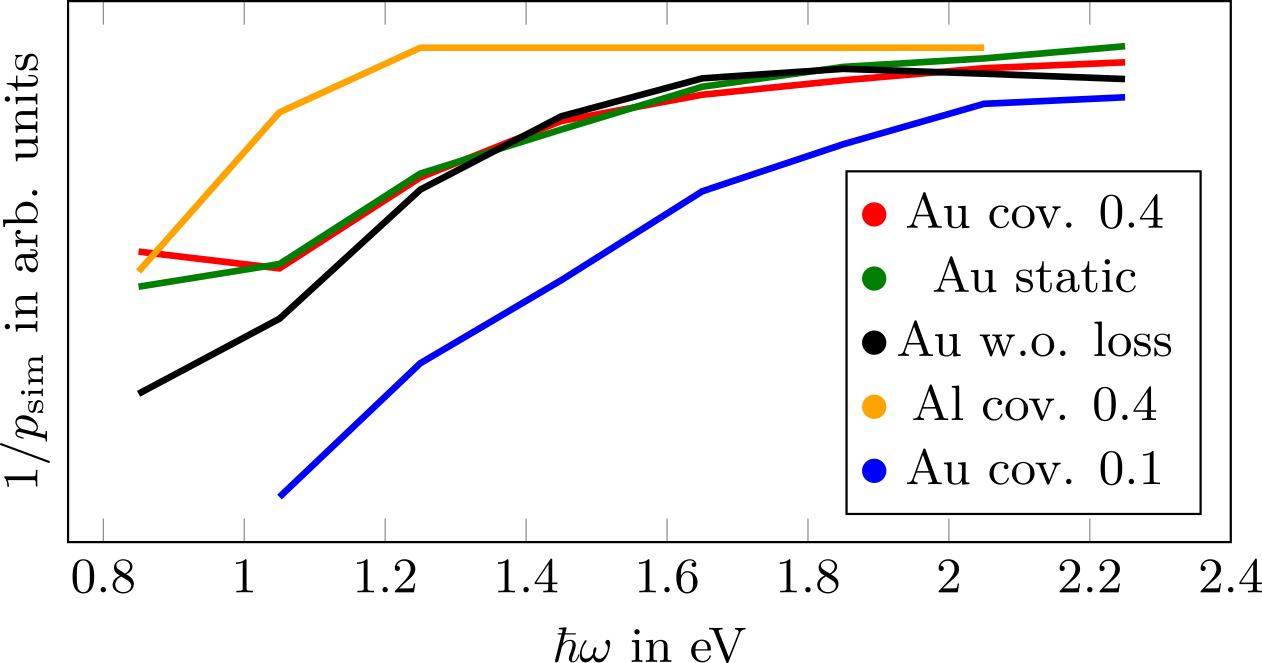}
		\caption{Simulated inverse participation number for different coverages (red and blue curves) and material (orange curve). The green and black graphs correspond to simulations of the quasistatic case (neglecting retardation effects) and loss free (no dielectric damping) material, both for gold and coverage of 40\,\%.}
		\label{fig:6}
	\end{figure}
	This resonance condition may be satisfied at low frequencies and filling factors as observed in Fig. \ref{fig:6},  where we show the results of a couple of simulations, showing the 2D Anderson nature of the localization, by:
	(I) a quasi-static simulation neglecting retardation effects,
	(II) a simulation assuming a loss free material (perfect conducting material with $\Im\left\{\varepsilon\right\}=0$),
	(III) a simulation of a web with much smaller coverage in comparison to the experimentally investigated one and (IV) a simulation of an aluminum web. Switching off retardation and loss doesn't lead to a significant change of the inverse participation number which proves that the localization is disorder driven.
	\renewcommand{\thefigure}{S4}
	\begin{figure}[!h]
		\centering
		\includegraphics[width=0.42\textwidth]{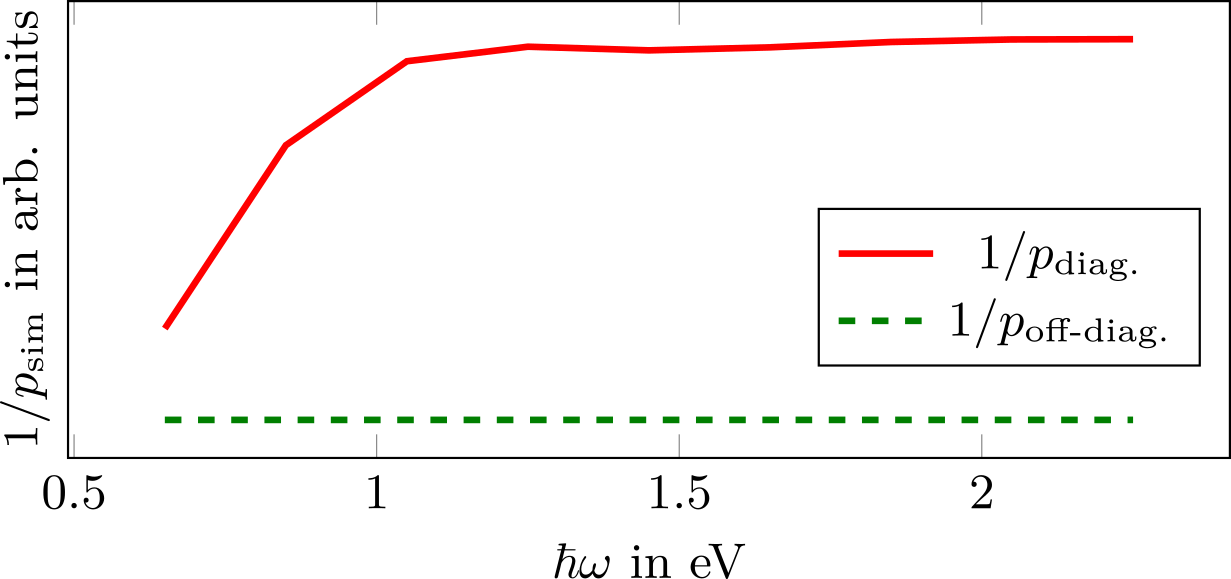}
		\caption{Effect of diagonal (NP geometry variation only) and off-diagonal (position randomization only) disorder on the inverse participation ratio.}
		\label{fig:8}
	\end{figure}
	Changing the coverage or material on the other hand, leads to a shift of the spectral threshold of vanishing hybridized LSP modes although the localization behavior in general remains (see Fig.\,\ref{fig:6}).
	 The particular form of the resonance problem in 2D (\ref{eq:SDCM2D}) and the associated eigenvalue equation (\ref{eq:eigSDCM2D}) allows separating the sources of disorder into diagonal ones ($\mathbf{D}$) and off-diagonal ones ($\mathbf{G}$) corresponding to the size and positional disorder of the NPs, respectively. Fig. \ref{fig:8} shows that in case of purely diagonal disorder the system doesn't show resonant modes (indicated by the dashed green line). The inverse participation number corresponding to off-diagonal disorder reproduces the general behavior of increasing localization at higher energies. Thus, the positional distribution of the holes determines the localized response in the 2D networks.
\end{appendix}	
	
\end{document}